\documentclass[a4paper,12pt]{article}
\usepackage{amsfonts}
\usepackage{latexsym}
\usepackage{amsmath}
\usepackage{amssymb}
\usepackage{amssymb}

\usepackage{slashed}
\usepackage{upgreek}
\usepackage{mathrsfs}
\usepackage{bbm}
\usepackage{enumerate}

\usepackage[toc,page]{appendix}

\usepackage{amsthm}
\usepackage[utf8]{inputenc}
\usepackage{graphicx}

\theoremstyle{remark}

\usepackage[utf8]{inputenc}
\usepackage[english]{babel}

\theoremstyle{definition}

\allowdisplaybreaks

\usepackage{relsize}
\usepackage{graphicx}
\usepackage{comment}

\renewcommand{\thefootnote}{\fnsymbol{footnote}}

\def\appendix#1{\addtocounter{section}{1}\setcounter{equation}{0}
	\renewcommand{\thesection}{\Alph{section}}
	\section*{Appendix \thesection\protect\indent \parbox[t]{11.15cm}{#1}}
	\addcontentsline{toc}{section}{Appendix \thesection\ \ \ #1}}

\newcommand{\pp}{=\kern-0.40em{\vert}}

\def\bbe{{\bf{e}}}

\font\mybb=msbm10 at 11pt

\def\bb#1{\hbox{\mybb#1}}

\def\bC {\bb{C}}

\newcommand{\bea}{\begin{eqnarray}}
	\newcommand{\eea}{\end{eqnarray}}

\usepackage[usenames,dvipsnames,svgnames,table]{xcolor}	
\usepackage[normalem]{ulem}				
\usepackage{microtype}
\usepackage[colorlinks, citecolor=blue, linkcolor=blue, urlcolor=Maroon, filecolor=Maroon, linktocpage=true]{hyperref} 

\usepackage{chngcntr}
\counterwithout{equation}{section}

\begin{document}

	\begin{center}
		\vspace*{-1.0cm}
		\begin{flushright}
		\end{flushright}

		
		\vspace{2.0cm} {\Large \bf On the geometry and quantum theory of regular and singular spinors} \\[.2cm]
		
		\vskip 2cm
		 G.\,  Papadopoulos
		\\
		\vskip .6cm


		\begin{small}
			\textit{Department of Mathematics
				\\
				King's College London
				\\
				Strand
				\\
				London WC2R 2LS, UK}
			\\*[.3cm]
			\texttt{george.papadopoulos@kcl.ac.uk}
		\end{small}
		\\*[.6cm]

	\end{center}

	\vskip 2.5 cm

	\begin{abstract}
\noindent
We relate the Lounesto classification of regular and singular spinors to the orbits of the $\mathrm{Spin}(3,1)$ group in the space of Dirac spinors.  We find that regular spinors are associated with the principal orbits   of the spin group while singular spinors are associated with special orbits whose isotropy group is $\bC$. We use this to clarify some aspects of the classical and quantum theory of spinors restricted to a  class in this classification. In particular, we show that the degrees of freedom of an ELKO field, which has been proposed as a candidate for dark matter, can be reexpressed as a  Dirac field preserving locality. Alternatively after introducing the ELKO dual,  it can  be re-interpreted as four anticommuting Lorentz scalar fields with internal symmetry the spin representation of the Lorentz group.  We also propose an interacting Lagrangian which can consistently describe all 6 classes of regular and singular spinors.

	\end{abstract}

	

	\newpage
	
	\renewcommand{\thefootnote}{\arabic{footnote}}

\section*{}

It is known for sometime that a Dirac spinor $\psi$ can be characterised by the properties of its form bilinears.   In four dimensions, these are spanned by the scalar $f=i\bar\psi\psi$, pseudo-scalar $f_5=\bar\psi\gamma_5\psi$, vector current $j_\mu=\bar\psi\gamma_\mu\psi$, axial current  $j_\mu^5=\bar\psi\gamma_\mu\gamma_5\psi$ and 2-form
$s_{\mu\nu}=\bar\psi\gamma_{\mu\nu}\psi$, where in a canonical Clifford algebra basis $\bar\psi$ is the Dirac conjugate of $\psi$, $\bar\psi=\psi^\dagger \gamma_0$, and $\psi^\dagger$ is the hermitian conjugate of $\psi$. These form bilinears are not independent as they are restricted by the Fierz identities. After a consideration of these identities, the spinors can be classified in   6 different classes \cite{loun}.  The spinors of the first three classes are characterised by the properties
\bea
(I):~f, f_5\not=0~,~~(II):~f\not=0, f_5=0~,~~~(III):~f=0, f_5\not=0~,~~~
\label{regular}
\eea
and are referred to as regular spinors.
While, the spinors of the remaining three classes are characterised by the properties  $f=f_5=0$ and
\bea
(IV):~s, j_5\not=0~,~~(V):~j_5=0, s\not=0~,~~~(VI):~j_5\not=0, s=0~,~~~
\label{singular}
\eea
and are referred to as singular spinors.   Note that $j$ cannot vanish  unless $\psi=0$ as $j^0=\psi^\dagger \psi$. These classes have been explored for various physics applications that include the construction of non-standard Wigner fields and proposals for dark matter, see e.g. \cite{dark, silva1, rocha1, silva2, rsv, rocha2} and review \cite{darkrev} including the references within.

One of the objectives of this paper is to relate the above classes to the orbit structure of the spin group\footnote{The spin group $\mathrm{Spin}(3,1)$ considered here is the double cover of the connected Lorentz group, i.e. the double cover of the group of proper and orthochronous Lorentz transformations.}, $\mathrm{Spin}(3,1)$, on the space of Dirac spinors and explore the above classes as solutions to the Dirac equation. We use these results to argue that the ELKO field, which has been proposed as a candidate for dark matter,  can be reexpressed as a Dirac field. Then, we comment on the consistency of theories with ELKO fields. An interacting theory that restricts the fields to $\psi$ take values in the above classes is also proposed.

An alternative way to characterise spinors is in terms of the orbits of the spin group, $\mathrm{Spin}(3,1)$, in the space of Dirac spinors, $\Delta$.  To do this  in four dimensions, it is convenient to represent the spinors in terms of forms and give a representative for each orbit, for more details see \cite{ugjggp}.  In this realisation of spinor representations, the space of Dirac spinors, $\Delta$, is identified with the space of forms of $\bC^2$, i.e. $\Delta=\Lambda^*(\bC^2)$.  Therefore, the most general Dirac spinor is $\psi=a_1\,1+ a_2\, e_{12}+ b_1\, e_1+ b_2\, e_2$, with $a_1,a_2,b_1,b_2\in \bC$, and $(1, e_1, e_2, e_{12})$ is a basis\footnote{This is a different basis from that of helicity states used in e.g. \cite{rsv} and it is more convenient for our considerations.} in $\Lambda^*(\bC^2)$ induced by the Hermitian basis $(e_1, e_2)$ of $\bC^2$, where $e_{12}=e_1\wedge e_2$ and $\wedge$ is the wedge product operation on the space of forms. The gamma matrices in this realisation of the spinor representation act as
\bea
\gamma_0= -e_2\wedge+ e_2\bar\wedge~,~~\gamma_1=e_1\wedge+ e_1\bar\wedge~,~~\gamma_2=e_2\wedge+ e_2\bar\wedge~,~~\gamma_3=i(e_1\wedge- e_1\bar\wedge)~,
\label{gamma}
\eea
where $\bar\wedge$ is the inner derivation operation, i.e. $e_i\bar\wedge 1=0, e_i\bar\wedge e_j=\delta_{ij}, e_i\bar\wedge e_{jk}=\delta_{ij} e_k-\delta_{ik} e_j$. The  positive (negative) chirality Weyl spinors, $\Delta^+$ ($\Delta^-$),  are spanned by the even (odd) degree forms and they are eigenspaces of the chirality operator $\gamma_5=i\gamma_{0123}$, i.e. as $\Delta=\Delta^+\oplus \Delta^-$, $\psi=\psi^++\psi^-$ with $\psi^+=a_1\,1+ a_2\, e_{12}$ and $\psi^-=b_1\, e_1+ b_2\, e_2$.

In what follows, we shall demonstrate that there are three types of orbits of $\mathrm{Spin}(3,1)$ on the space of Dirac spinors represented by
\bea
(i):~\psi=0~;~~(ii)~\psi=1+ b\, e_2~,~~b\in\bC-\{0\}~;~~(iii)~\psi= a_0\, 1+ a_1\, e_1~,~~~(a_0, a_1)\in \bC\mathrm{P}^1~,
\label{orb}
\eea
where the equivalence relation on the components $(a_0, a_1)$ is defined below.
Clearly, one of the orbits of the spin group is the trivial orbit $\psi=0$. If $\psi\not=0$, there are several cases to consider. If both $\psi^+, \psi^-\not=0$, then it is always possible to arrange such that $\psi^+=1$ as the action of  $\mathrm{Spin}(3,1)$ on $\Delta^+$ in the $(1, e_{12})$ basis can be identified with that of $SL(2,\bC)$ on $\bC^2$. The subgroup of $SL(2,\bC)$ generated by $(\gamma_1 (\gamma_0+\gamma_2), \gamma_3(\gamma_0+\gamma_2))$ that leaves the  spinor $1$ invariant  is $\bC$ embedded in $SL(2,\bC)$ as
\bea
\begin{pmatrix}
1 & \lambda
\\
0& 1
\end{pmatrix}
\eea
where $\lambda\in \bC$. This can be used to find a representative for the orbits of $\mathrm{Spin}(3,1)$ in $\Delta^-$.  It can be easily seen that there are two possibilities. One is that the component of $\psi^-$ along $e_2$ is non-zero, i.e. $b_2\not=0$.  In such a case, after using a $\bC$ transformation, one can always set $b_1=0$. As a result, a representative of this class of orbits is
\bea
\psi=1+ b e_2~,~~b\in\bC-\{0\}~.
\label{typei}
\eea
The isotropy group of this spinor in $\mathrm{Spin}(3,1)$ is the identity and so every such orbit of  $\mathrm{Spin}(3,1)$ in $\Delta$ can be identified with $\mathrm{Spin}(3,1)$ for each $b\in\bC-\{0\}$. Therefore these orbits are of co-dimension two and have the highest dimension of any other orbit of $\mathrm{Spin}(3,1)$  in $\Delta$.  As a result, these are the principal orbits of the group and their union is dense in $\Delta$.

The second possibility is that $b_2=0$. If this is the case, $\psi^-$ is invariant under the action of the $\bC$ subgroup of $SL(2,\bC)$.  As a result, a representative for this class of orbits is
\bea
\psi=1+ a e_1~,~~a\in\bC-\{0\}~.
\label{typeii0}
\eea
As the isotropy group of these spinors is $\bC$, each of these orbits can be identified with $SL(2,\bC)/\bC$.

The next case to consider is whenever either $\psi^+$ or $\psi^-$ vanishes. If $\psi^-=0$, then $\psi=1$ up to an $SL(2,\bC)$ transformation.  The orbit can be identified with $SL(2,\bC)/\bC$.  Clearly, this is a special case of the orbits represented by  the spinors (\ref{typeii0}) provided that one allows $a$ to vanish.
If $\psi^+=0$, one can use an $\mathrm{Spin}(3,1)$ transformation to set $\psi= e_1$ with isotropy group $\bC$.

It turns out that the orbits represented with the spinors (\ref{typeii0}) and those with either $\psi^+=0$ or $\psi^-=0$ can be described together as follows.
Consider the spinor $\psi= a_0 1+ a_1 e_1$ and observe that it is in the same $\mathrm{Spin}(3,1)$ orbit as the spinor $\psi'= \mu a_0 1+ \bar\mu a_1 e_1$, $\mu\in \bC-\{0\}$, where $\bar\mu$ is the complex conjugate of $\mu$. This transformation is generated by $\gamma_{02}$ and $\gamma_{13}$. As a result $\psi=1+a e_1$ and $\psi'= a' 1+ e_1$ represent the same orbit provided that for $a, a'\not=0$, $a'=\bar a^{-1}$.  Thus all these orbits can be represented together with the spinors
\bea
 \psi= a_0 1+ a_1 e_1~,
 \label{typeiia}
  \eea
  provided that one identifies the representatives with components $(a_0, a_1)\in \bC^2-\{0\}$ with those with components $(\mu a_0, \bar\mu a_1)$, $\mu\in\bC-\{0\}$.  Even though this identification is not the standard relation defining $\bC\mathrm{P}^1$, the independent pairs of components parameterise $\bC\mathrm{P}^1$.
Therefore, there is a 2-sphere parameter family of $SL(2,\bC)/\bC$ orbits with the north and south poles representing the orbits of $\mathrm{Spin}(3,1)$ on the space of chiral and anti-chiral Weyl spinors, respectively.  Clearly, these orbits are of co-dimension four in $\Delta$.

To relate the characterisation of spinors in terms of the orbits of $\mathrm{Spin}(3,1)$ in $\Delta$ with the Lounesto's classification  in terms of form bilinears, it suffices to compute the form bilinears of the spinors that represent the orbits given in (\ref{orb}). This is because both characterisations are Lorentz covariant and that in terms of the form bilinears relies on their vanishing conditions. Therefore, if a form bilinear vanishes in a Lorentz frame, e.g. the Lorentz frame chosen to describe the representatives
of the orbits as in (\ref{orb}),  it will vanish in every Lorentz frame.  The form bilinears of the spinors that represent  the principal orbits (\ref{typei}) are
\bea
&&f=i(\bar b-b)~,~~~f_5=\bar b+ b~,~~~j=(1+b \bar b)\, \bbe^0+ (-1+\bar b b)\, \bbe^2~,~~~
\cr
&&
j_5=(1-\bar b b)\bbe^0- (1+\bar b b)\, \bbe^2~,~~~
 s=(b+\bar b)\, \bbe^0\wedge \bbe^2+i (\bar b-b)\, \bbe^1\wedge \bbe^3~,
\eea
where $(\bbe^0, \bbe^1, \bbe^2, \bbe^3)$ is a pseudo-orthonormal frame, i.e. the Minkowski metric is expressed as $\eta=-(\bbe^0)^2+ (\bbe^1)^2+(\bbe^2)^2+(\bbe^3)^2$,  and so in Cartesian coordinates $\bbe^\mu=dx^\mu$.
To explain this computation consider that of $j_5$ in more detail. In particular, one has that
\bea
&&j_5\equiv\bar\psi\gamma_\mu \gamma_5 \psi \,\bbe^\mu \equiv \langle \gamma_0 \psi, \gamma_\mu\gamma_5\psi \rangle\, \bbe^\mu=\langle -e_2+ b 1, \gamma_\mu (1- b e_2)\rangle\, \bbe^\mu
\cr
&&\quad= \langle -e_2+ b 1, \gamma_0 (1- b e_2)\rangle\, \bbe^0+ \langle -e_2+ b 1, \gamma_2 (1- b e_2)\rangle\, \bbe^2
\cr
&&\quad =\langle-e_2+ b 1, -e_2-b 1\rangle\, \bbe^0
+\langle -e_2+ b1, e_2- b1\rangle\, \bbe^2
\cr
&&\quad
= (1-b\bar b)\, \bbe^0-(1+b\bar b)\, \bbe^2~,
\eea
where $\langle\cdot, \cdot\rangle$ is the standard Hermitian inner product, $\psi=1+ be_2$ and similarly for the rest of the form bilinears.

Moreover after a similar computation, the form bilinears  of the spinors (\ref{typeiia}) that represent the remaining orbits are
\bea
&&f=f_5=0~,~~~j=(\bar a_0 a_0+ \bar a_1 a_1)\, \bbe^0- (\bar a_0 a_0+ \bar a_1 a_1)\, \bbe^2~,~~~
\cr
&&
j_5=(\bar a_0 a_0- \bar a_1 a_1)\, \bbe^0- (\bar a_0 a_0- \bar a_1 a_1)\, \bbe^2~,~~~
\cr
&&
s=(\bar a_1 a_0+\bar a_0 a_1)\, \bbe^0\wedge \bbe^1+i (\bar a_1 a_0-\bar a_0 a_1)\, \bbe^0\wedge \bbe^3
\cr
&& \qquad +( \bar a_1  a_0
 +\bar a_0 a_1)\, \bbe^1\wedge \bbe^2
-i (\bar a_1 a_0-\bar a_0 a_1)\, \bbe^2\wedge \bbe^3~.
\eea
The spinors (\ref{typei}) representing the principal orbits of $SL(2,\bC)$ in $\Delta$, and therefore all spinors that lie on the principal orbits, belong to the form bilinears class (I) in (\ref{regular}). The  (II) and (III) classes in (\ref{regular}), which depend on whether the complex parameter $b$ is imaginary or real, respectively, correspond to a subclass of principal orbits.

 The  spinors (\ref{typeiia}) that represent the remaining orbits belong to the form bilinear classes (IV), (V) or (VI) depending on the choice of parameters $(a_0, a_1)$.  For a generic choice, they belong to class (IV),  while for $\bar a_0 a_0- \bar a_1 a_1=0$ in the class (V) and for $\bar a_1 a_0=0$ in the class (VI).

 Clearly for the spinors that represent the class (V), one has $a_0, a_1\not=0$. Setting $a_0=1$, one finds that $a_1=e^{i\theta}$ and so the orbits of the spin group that belong in the class (V) are parameterised by the angle $\theta$. For each $\theta$, the  spinor that represents the class (V) is real  with respect to the charge conjugation operator $C(\theta)=- e^{i\theta} \gamma_{012}*$,  i.e. $\psi$ is Majorana with respect to $C(\theta)$, where $*$ is the standard complex conjugation operation, $*\mu=\bar\mu$, $\mu\in \bC$. Therefore, the class (V) contains the Majorana spinors for {\sl all} possible choices of a reality condition.

 The representative spinors of class (VI)  are  Weyl spinors one for each chirality. Denoting with, (V$_C$), the spinors of class (V) that are real with respect to a fixed $C$,  both classes (V$_C$)  and (VI) span vector spaces that are preserved by the action of the spin group. This is the case for the class (V$_C$) spinors as $C$  commutes with the action of $\mathrm{Spin}(3,1)$ and $\mathrm{Spin}(3,1)$ acts transitively on the space of all non-vanishing Majorana spinors.

Note that for the spinors that lie in the principal orbits, the bilinears $f$ and $f_5$ determine the spinor in the Lorentz frame that the spinor takes the canonical form as in (\ref{orb}). Similarly, the bilinear $s$ determines the spinors that lie in the remaining orbits. This is  an example of an inversion theorem \cite{tak} that describes how the spinors can be reconstructed  from their form bilinears.

For applications  to physics, one can either impose restrictions on the form bilinears of a spinor in momentum space or in position space. In the former case, which has been mostly explored in the literature, the solutions of the Dirac equation with  mass term lie in the classes (I) and (II), i.e. they are  elements of the principal  $SL(2,\bC)$ orbits.  To see this, one can substitute the representative $u(p)=1+b e_2$ into Dirac equation $(i \slashed{p}+m) u(p)=0$, where $m$ is the rest mass $m>0$. Notice that this is equivalent to solving the Dirac equation in the Lorentz frame that the solution takes the form $1+b e_2$.   Then, one finds that there is a solution provided that
\bea
p^1=p^3=0~,~~~b=-i m^{-1} (p^0+p^2)~,~~~ \mathrm{and}~~(p^0)^2=(p_2)^2+m^2~,
\eea
 which is interpreted as a spinor with mass $m$ propagating in the $x^2$ direction. As $b$ is purely imaginary, $f_5=0$,   all  solutions of the Dirac equation do not lie in the class (III).  Furthermore, if in addition $f=0$ and so $b=0$, the solution becomes that of a massless Dirac fermion, $m=0$, propagating in the $x^2$ direction with energy $p^0=-p^2$.

The same applies for the spinors in the class (V$_C$) with the understanding that in momentum space the Dirac equation is solved for  complex spinors and the Majorana solution in position space is constructed by taking the real part of the complex solution with respect to $C$. However, as we shall demonstrate below, the Dirac equation in momentum space does not admit solutions that belong in the class (V$_C$) unless they are tachyonic.    The spinors of class (VI) are the usual chiral solutions of the Dirac equation with $m=0$.

To get an insight into the class of  (IV) spinors as solutions of the Dirac equation, let us solve $(i \slashed{p}+m) u(p)=0$ for $u(p)=1+ a e_1$. If $a=0$, there is a solution provided that $m=0$ and $p^0=-p^2$. This is the usual solution for a chiral fermion moving in the $x^2$ direction. For $a\not=0$, the Dirac equation gives $p^0=-p^2$, $p^1=2^{-1} m i ( a^{-1}+a)$ and $p^3=-2^{-1} m ( a^{-1}-a)$. Imposing the reality condition on all  components of the momentum leads to an inconsistency. However, notice that if either $a^2=-1$ or $a^2=1$ and $m$ is purely imaginary, then there is a consistent solution but such a solution is a tachyon -- for $a^2=\pm1$ each solution in momentum space is a Majorana spinor with respect to a reality condition $C$ but $C$ depends on the solution.    Furthermore, as the spinors of class (IV) obey the quadratic equations $f=f_5=0$ do not span a vector space. Nevertheless, one can take such spinors, related by a Lorentz transformation, and try to find solutions of the Dirac equation. In such a case, one can show that Lorentz transformation related spinors of  class (IV) span the whole $\Delta$. Indeed,  notice that $1+a e_1$, $i(1-a e_1)$, $e_{12}-a e_2$ and $i(e_{12}+a e_2)$, which are related by Lorentz transformations and each belongs to the class (IV) of spinors, $a\not=0$,  is a basis in $\Delta$. A consequence of this is that their linear superposition is not always a spinor in the (IV) class. Despite the differences mentioned above in the behaviour of class (IV) and class (VI) spinors as solutions to the Dirac equation, they belong to the same type of orbit of the $\mathrm{Spin}(3,1)$ group and they are continuously connected. So  the differences in their behaviour\footnote{This is a reminiscent of the debate about the nature of neutrinos. In particular, the parts of the debating regarding the question on whether they are Majorana spinors or not, after originally thought to be chiral, and on whether they exhibit tachyonic behaviour.} remains  a puzzle.

As an application of the formalism above, let us consider the construction\footnote{The application of the  Lounesto classification to ELKO field was proposed in \cite{rocha}.}  of ELKO field \cite{dark}.  To make connection with the results in the literature, from now on,  we use a spacetime metric with signature mostly minus instead of that with signature most plus  employed so far. This is easily done by multiplying the gamma matrices $\gamma$ in (\ref{gamma}) with an $i$, i.e. set $\Gamma_\mu=i\gamma_\mu$. The complex conjugation operation is again given by $C(\theta)=- e^{i\theta} \Gamma_{012}*$ and satisfies $C^2=1$ but now it anti-commutes with the new gamma matrices $\Gamma_\mu$.  Following the same notation as in \cite{darkrev}, an ELKO basis of spinors can be explicitly constructed as
\bea
&&\lambda^S_+=\sqrt{m}(1+e_1)~,~~~    \lambda^S_-=\sqrt{m} (e_2-e_{12})~,~~~
\cr
&&\lambda^A_+=\sqrt{m}(1-e_1)~,~~\lambda^A_-= -\sqrt{m} (e_2+e_{12})~,
\label{elko}
\eea
where  the spinors $\lambda^S_\alpha$ and $\lambda^A_\alpha$  are real with  respect to $C=-i\Gamma_{012}*=-\gamma_{012}*$ and  $-C$, respectively, i.e. they satisfy equation (22) in \cite{darkrev} as required\footnote{The basis for ELKO spinors (\ref{elko}) differs from those traditionally chosen in the literature. However, it satisfies the key equation required, i.e. eqn (22) in \cite{darkrev}, and so it is related to those in the literature up to a Lorentz transformation and a choice of reality condition.  These choices do not affect the physical properties of the system.}.
One can introduce a basis for the $\rho^S$ and $\rho^A$ spinors in a similar way. For this choice of $C$ as a reality condition, the basis (\ref{elko}) is essentially unique up to a Lorentz transformation. The same conclusion holds for any other choice of $C$.  As expected, this basis exhibits the same subtleties regarding their spin sums as others chosen in the literature.
Notice that
\bea
\slashed{p}\lambda_+^S=i m \lambda_-^S~,~~~\slashed{p}\lambda_-^S=-i m \lambda_+^S~, ~~~\slashed{p}\lambda_+^A=-i m \lambda_-^A~, ~~~\slashed{p}\lambda_-^A=i m \lambda_+^A~,
\label{lsa}
\eea
 in the Lorentz frame that $p^1=p^3=0$ and $-p^0+p^2=m$.   Then, these can be boosted with a Lorentz transformation to hold for any momentum $k$. A consequence of (\ref{lsa}) is that $\lambda^S$ and $\lambda^A$ satisfy $(k^2-m^2) \lambda_\alpha^{S,A}=0$, i.e. $k$ is on-shell. However, it should be noted that the converse is not true as $(k^2-m^2) \lambda_\alpha^{S,A}=0$  does not imply the conditions (\ref{lsa}) with $p=k$. Instead, one  observes that
 \bea
 (\slashed{k}\mp m) (\lambda_+^S\pm i \lambda_-^S)=0~,~~~(\slashed{k}\pm m) (\lambda_+^A\pm i \lambda_-^A)=0~.~~~
 \eea
 Furthermore, one has that
 \bea
 (\slashed{k}\pm m) \Gamma_5(\lambda_+^S\pm i \lambda_-^S)=0~,~~~(\slashed{k}\mp m) \Gamma_5(\lambda_+^A\pm i \lambda_-^A)=0~,~~~
 \eea
 where $\lambda_+^A+ i \lambda_-^A$, $\Gamma_5(\lambda_+^A- i \lambda_-^A)$ , $\lambda_+^S+ i \lambda_-^S$ and $\Gamma_5 (\lambda_+^S- i \lambda_-^S)$ are linearly independent and $\Gamma_5=\gamma_5$.

Postponing the discussion about spin sums  for later, one can proceed with the introduction of ELKO field\footnote{Even though $\lambda^{S,A}$ satisfy reality conditions, the field $\psi_e$ is complex. Moreover, the reliance of the construction on the momentum space representation is a limiting factor. For example,  such an expression for the fields does not exist on curved spacetimes and so one cannot impose the reality conditions on $\lambda_\alpha^{A,S}$  as these cannot be defined. What remains and can be used are the Klein-Gordon type of field equations, see eqn (\ref{KGe}),  that the field satisfies that admit a generalisation  to curved spacetimes.} as
\bea
\psi_e(x)=\int d\mu(\vec k) \big(\sum_\alpha a^\dagger_\alpha(k) \lambda^S_\alpha  e^{i kx}+ \sum_\alpha b_\alpha(k) \lambda^A_\alpha  e^{-i kx} \big)~,
\label{phix}
\eea
in the usual way, where $d\mu$ is an appropriate measure and $(b^\dagger_\alpha, a^\dagger_\alpha)$ and $(b_\alpha, a_\alpha)$ are creation and annihilation operators, respectively. Departing from the standard construction, one can also introduce the field
\bea
\chi_e(x)= \int d\mu(\vec k) \sum_{\alpha, \beta} \epsilon^{\alpha\beta}\big( a^\dagger_\alpha(k) \lambda^S_\beta  e^{i kx}+ b_\alpha(k) \lambda^A_\beta  e^{-i kx} \big)~,
\eea
where $\epsilon^{\alpha\beta}=-\epsilon^{\beta\alpha}$ and $\epsilon^{+-}=1$.  As a consequence of (\ref{lsa}), one finds that
\bea
\slashed{\partial}\psi_e(x)=-m \chi_e(x)~,~~~\slashed{\partial}\chi_e(x)=m \psi_e(x)~,
\eea
and so
\bea
(i\slashed{\partial}+m) (\psi_e+i\chi_e)=0~,~~~(i\slashed{\partial}-m) (\psi_e-i\chi_e)=0~.
\eea
Therefore,  $\psi_e+ i\chi_e$ solves the Dirac equation but it is not a standard Dirac field.  This is because
 \bea
 \psi_e+i\chi_e=\int d\mu(\vec k) \big((a^\dagger_+-i a^\dagger_-) (\lambda_+^S+i \lambda_-^S) e^{i kx}+ (b_+-i b_-) (\lambda_+^A+i \lambda_-^A) e^{-i kx}\big)~,
 \label{xyz}
 \eea
 and so  it depends on a single creation operator $\tilde a^\dagger= a^\dagger_+-i a^\dagger_-$ and a single annihilation operator $\tilde b=b_+-i b_-$.  As a result, it describes a particle  anti-particle pair with  only one of the two spin directions expected for massive fermions as unitary representations of the Poincar\'e group. Similarly, $\psi_e- i\chi_e$ satisfies the Dirac equation after setting $m$ to $-m$ and describes a particle anti-particle pair with a single spin direction  each. This leads to subtleties in the definition of the spin sums and the Feynman propagator of the ELKO fields, see e.g. \cite{darkrev} and references therein as well as the discussion below.

However, the two fields can be reassembled  to a Dirac field as
\bea
\Psi=\psi_e+ i\chi_e+ \Gamma_5 (\psi_e- i\chi_e)~,
\label{direlko}
\eea
with each particle and the anti-particle  exhibiting two independent spin directions as expected.  These are given by   $\lambda_+^A+ i \lambda_-^A$ and $\Gamma_5(\lambda_+^A- i \lambda_-^A)$ for the particle, and by  $\lambda_+^S+ i \lambda_-^S$ and $\Gamma_5 (\lambda_+^S- i \lambda_-^S)$ for the antiparticle.
 The expression of $\Psi$ in terms of $\psi_e$   is local. The spin sums for the  field $\Psi$ are standard\footnote{Furthermore under parity transformations $(x^0, \vec x)\rightarrow (x^0, -\vec x)$, $\Psi$ transforms in the standard way, i.e. in the basis used here $\Psi'(x^0, -\vec x)=\Gamma_0 \Psi(x^0, \vec x)$.} and quantisation leads to fermion statistics   and to the standard Feynman propagator for the Dirac fermion.

  The above results raise the question about the consistency of the proposed quantisation  of theories with ELKO fields, see also \cite{romero}. The proposal is that the ELKO field is a fermion and its  propagator is  the standard  Feynman propagator of a scalar field, see also further comment below.    However, if a theory with an  ELKO field has suitable couplings such that it can be reexpressed as a theory of   a Dirac field, then  consistency  will require that the ELKO field is a fermion and its Feynman propagator is related to the standard Feynman propagator of a Dirac fermion instead of that of a scalar field.   Putting it in a different way, let us consider a theory with a Dirac field which is rewritten in terms of an ELKO field using (\ref{direlko}), after focusing on the ELKO field, one may come to the wrong conclusion that the theory is inconsistent because the ELKO field does not have the correct kinetic term.

 Next, let us focus on the dynamics of the ELKO field as described in the literature, see e.g. review \cite{darkrev}  and references therein. One of the features of producing a consistent local theory for the ELKO field is the introduction of a conjugate, the ELKO conjugate or dual, $\breve{\psi}_e$,  of $\psi_e$. The reason for not using the Dirac conjugate $\bar\psi_e$ for the ELKO field $\psi_e$ is that the spin sums $\sum_\alpha \lambda^{S,A}_\alpha\otimes \bar \lambda^{S,A}_\alpha$ are degenerate\footnote{Clearly, the degeneracy in the spin sums above is closely related to the observation made below (\ref{xyz}) that the field $\psi_e+i\chi_e$ does not have sufficient degrees of freedom to be interpreted as a Dirac field.} and so the Feynman propagator of the $\psi_e$ field has zero modes. According to the literature this leads to a non-local theory.
 To avoid this outcome,
   the ELKO conjugate $\breve{\psi}_e$ has been proposed instead, which in momentum space, is related to $\psi_e$ via an expression that depends non-trivially on the momentum\footnote{This is already a limitation in the theory as such a conjugate cannot be defined on curved spacetimes that the quantum theory does not have a momentum representation.} $k$, see e.g. \cite{darkrev}. This means that in position space, the conjugate $\breve{\psi}_e$, which also defines an inner product on the space of ELKO fields, is related to ${\psi}_e$ via a possible non-local  operator.  In addition,  the procedure of producing the ELKO conjugate and  the treatment of  spin sums for ELKO fields involves the construction of an inverse to a  degenerate matrix through a procedure that involves a limit.  To my knowledge, there is not an explicit expression in the literature that  gives the ELKO conjugate  $\breve{\psi}_e$ in terms of $\breve{\psi}_e$ for  any momentum $k$ and the corresponding expression of  the ELKO dual $\breve{\psi}_e$ in position space.  Assuming that all these operations can be consistently defined, let us focus on the final result that one arrives after introducing the ELKO conjugate and spins sums as advocated in the literature. The claim is that the ELKO field  has mass dimension one, fermionic statistics, transforms under the spin representation of the Lorentz group  and with Feynman propagator
 ${\mathbf 1}(p^2-m^2+i\epsilon)^{-1}$, see e.g. \cite{darkrev} equations (63) to (68). In position space the Lagrangian of an ELKO field is
 \bea
 {\cal L}=\breve{\psi}_e (\partial^2+m^2)\psi_e~,
 \label{KGe}
 \eea
and the field satisfies that Klein-Gordon (KG) equation, where $\breve\psi_e$ is the ELKO  conjugate of $\psi_e$. Apart from the profound differences in the physical interpretation of the fields that satisfy the Dirac and KG equations, even if the latter transform under the spin representation of the Lorentz group, there is an additional one which is not as emphasised. To explain, covariance  of the Dirac equation under Lorentz transformations demands that the Dirac fields transform as
\bea
\psi'(x')= S(\Lambda) \psi(x)~,
\eea
where $x'^\mu=\Lambda^\mu{}_\nu x^\nu$, $\Lambda$ is a Lorentz transformation  and $S(\Lambda)$ is in the spinor representation that depends on $\Lambda$. Altogether, $\Lambda$ and $S(\Lambda)$ are required for the Dirac action to be invariant under the {\sl spacetime}\footnote{Here, the terms internal and spacetime symmetries are used as in the context of the Noether's theorem.} symmetries generated by the Lorentz group.

This is not the case for the KG action (\ref{KGe}). The KG action (\ref{KGe}) is separately invariant under the {\sl spacetime} symmetries $\psi_e'(x')=\psi_e(x)$, where again $x'^\mu=\Lambda^\mu{}_\nu x^\nu$,  and the {\sl internal} symmetries $\psi_e'(x)= S\, \psi_e(x)$, where $S$ is a spin transformation -- notice that $x$ is not transformed in the second transformation and $S$ does not depend on $\Lambda$. Thus, one can conclude that the ELKO field consists of  four (anticommuting)  Lorentz {\sl scalar} fields that possibly transform under the {\sl internal} symmetry\footnote{Field theories with internal symmetries generated by non-compact groups have potential pathologies, like for example the Hamiltonians are not bounded from below and the kinetic terms of fields carry the wrong sign. This is because for such groups the invariant inner products are indefinite.} associated to the spin representation of the Lorentz group.

 For applications to dark matter, it is clear from the above analysis that the ELKO field can be linearly reexpressed as a Dirac field in a way that preserves locality. Therefore, it is possible to envisage theories for which the ELKO field can minimally couple to other fields,  like for example the electromagnetic field, as part of the standard coupling of the Dirac fermion.  This allows the potential coupling of the ELKO field to other fields and, therefore,  a much more detailed analysis is required in order to be considered as a dark matter candidate. Alternatively taking into account the ELKO conjugate,   the application of the ELKO field to dark matter is as that of  four anti-commuting scalar fields.  In particular, they can couple as doublets to standard model fields. This will break the internal symmetry generated by the spin representation of the Lorentz group but not the Lorentz invariance of the theory. Again, a more careful analysis is needed for the ELKO field to be considered as a dark matter candidate.

An alternative way to explore applications is to impose the restrictions on the form bilinears of a spinor in position space.  These can be imposed by a scalar, $\phi$, pseudo-scalar, $\phi_5$, 1-forms, $A$, and, $B$, and 2-form $G$ Lagrange multipliers. The associated Lagrangian can be written as
\bea
{\cal L}=\bar\psi (i\slashed{D}+m) \psi+ \phi (\bar\psi \psi- \alpha)+ \phi_5 (\bar\psi\gamma_5 \psi- \beta)+ \bar\psi \slashed {B} \gamma_5\psi+{i\over2}
\bar\psi \slashed {G} \psi~,
\label{lag}
\eea
where $D_\mu=\partial_\mu-i e A_\mu$ and $\alpha$ and $\beta$ are couplings which can be identified with the classical value of the scalar bilinears after imposing the
field equation of $\phi$ and $\phi_5$, respectively.  Clearly, this is an interacting theory that can be easily generalised to curved spacetimes. Also, it does not rely on the momentum space representation of field theory that does not exist on curved spacetimes. The regular spinors without further restrictions are described by the above Lagrangian after setting all the Lagrange multipliers to zero, $A=B=\phi=\phi_5=G=0$. The theory is that of a free Dirac field. To describe the classes (II) and (III), one has to set $A=B=G=0$ and either $\phi=0$ and $\beta=0$  or $\phi_5=0$ and $\alpha=0$, respectively.  To describe the class (IV) of singular spinors, one has to set $A=B=G=0$. While to describe the classes (V) and (VI) of singular spinors, one has to set either $A=B=0$ or $A=G=0$, respectively.

The quantum theory of systems described by the Lagrangian  (\ref{lag}) has been considered before in the literature \cite{eguchi, ebert1, ebert2, gaillard, h1, stergiou} in a variety of contexts. This includes the theory of mesons as described by the Nambu-Jona-Lasino action after performing a Hubbard-Stratonovich transformation.  Typically, the focus has been to integrate over the  fermionic field $\psi$, which is  taken as anticommuting. This produces an effective theory for the Lagrange multipliers. In particular,  the effective action of the theory after integrating $\psi$ can be expressed as
\bea
\Gamma=-i\,\mathrm{Tr}\log \big(i\slashed{D}+m+\phi +\phi_5 \gamma_5 +\slashed{B}\gamma_5+{i\over2}\slashed {G} \big)-\int d^4x \big(\alpha \phi+\beta \phi_5\big)~.
\eea
This is computed by considering diagrams whose  external lines are only the Lagrange multipliers and the fermion $\psi$ propagates  in the loops.
Further analysis reveals that in an number of occasions, the above effective action produces in a low energy limit a dynamical theory for the Lagrange multipliers \cite{gaillard, h1, stergiou, ellis, ah, smith}; for a recent careful derivation see \cite{stergiou}.
However, the appropriate limit to consider here is that for which the Lagrange multipliers are integrated over and the only external lines are those of the fermionic field $\psi$.  We shall not present the full analysis here. Nevertheless, some cases are straightforward.  For example, if the only non-vanishing Lagrange multiplier is $\phi$ and $\alpha=0$, then integrating over $\phi$ induces a delta function, $\delta(\prod_x f(x))$, which sets $f$ to zero at every spacetime point. As a result the theory describes a massless free Dirac fermion. This is consistent with the result we have derived after investigating the classical solutions of the Dirac equation for which the form bilinear $f$ vanishes.


\section*{Acknowledgments}
I would like to thank Cheng-Yang Lee for correspondence.






\begin{thebibliography}{99}

\bibitem{loun}
P.~Lounesto,
``Clifford Algebras and Spinors''
Lecture Note Series \textbf{286}, CUP (2001), page 286
doi:10.1088/1126-6708/2003/04/007
[arXiv:hep-th/0301161 [hep-th]].


\bibitem{dark}
D.~V.~Ahluwalia and D.~Grumiller,
``Dark matter: A Spin one half fermion field with mass dimension one?,''
Phys. Rev. D \textbf{72} (2005), 067701
doi:10.1103/PhysRevD.72.067701
[arXiv:hep-th/0410192 [hep-th]].

\bibitem{silva1}
J.~M.~Hoff da Silva and R.~da Rocha,
``Unfolding Physics from the Algebraic Classification of Spinor Fields,''
Phys. Lett. B \textbf{718} (2013), 1519-1523
doi:10.1016/j.physletb.2012.12.026
[arXiv:1212.2406 [hep-th]].

\bibitem{rocha1}
R.~Ab\l{}amowicz, I.~Gon\c{c}alves and R.~da Rocha,
``Bilinear Covariants and Spinor Fields Duality in Quantum Clifford Algebras,''
J. Math. Phys. \textbf{55} (2014), 103501
doi:10.1063/1.4896395
[arXiv:1409.4550 [math-ph]].

\bibitem{silva2}
J.~M.~Hoff da Silva and R.~T.~Cavalcanti,
``Further investigation of mass dimension one fermionic duals,''
Phys. Lett. A \textbf{383} (2019) no.15, 1683-1688
doi:10.1016/j.physleta.2019.02.041
[arXiv:1904.03999 [physics.gen-ph]].

\bibitem{rsv}
R.~J.~Bueno Rogerio, J.~M.~Hoff da Silva and C.~H.~Coronado Villalobos,
``Regular spinors and fermionic fields,''
Phys. Lett. A \textbf{402} (2021), 127368
doi:10.1016/j.physleta.2021.127368
[arXiv:2010.08597 [hep-th]].





\bibitem{rocha2}
G.~B.~de Gracia, A.~A.~Nogueira and R.~da Rocha,
``Fermionic dark matter-photon quantum interaction: A mechanism for darkness,''
Nucl. Phys. B \textbf{992} (2023), 116227
doi:10.1016/j.nuclphysb.2023.116227
[arXiv:2302.06948 [hep-ph]].



\bibitem{darkrev}
D.~V.~Ahluwalia, J.~M.~H.~da Silva, C.~Y.~Lee, Y.~X.~Liu, S.~H.~Pereira and M.~M.~Sorkhi,
``Mass dimension one fermions: Constructing darkness,''
Phys. Rept. \textbf{967} (2022), 1-43
doi:10.1016/j.physrep.2022.04.003
[arXiv:2205.04754 [hep-ph]].



\bibitem{ugjggp}
U.~Gran, J.~Gutowski and G.~Papadopoulos,
``Geometry of all supersymmetric four-dimensional N = 1 supergravity backgrounds,''
JHEP \textbf{06} (2008), 102
doi:10.1088/1126-6708/2008/06/102
[arXiv:0802.1779 [hep-th]].

\bibitem{tak}
Y.~Takahashi,
``Reconstruction of Spinor From Fierz Identities,''
Phys. Rev. D \textbf{26} (1982), 2169
doi:10.1103/PhysRevD.26.2169

\bibitem{rocha}
R.~da Rocha and W.~A.~Rodrigues, Jr.,
``Where are ELKO spinor fields in Lounesto spinor field classification?,''
Mod. Phys. Lett. A \textbf{21} (2006), 65-74
doi:10.1142/S0217732306018482
[arXiv:math-ph/0506075 [math-ph]].

\bibitem{romero}
R.~Romero,
``Elko spinors revised,''
Rev. Mex. Fis. \textbf{69} (2023) no.2, 020201
doi:10.31349/RevMexFis.69.020201
[arXiv:2207.08334 [hep-th]].

\bibitem{eguchi}
T.~Eguchi,
``A New Approach to Collective Phenomena in Superconductivity Models,''
Phys. Rev. D \textbf{14} (1976), 2755
doi:10.1103/PhysRevD.14.2755


\bibitem{ebert1}
D.~Ebert and M.~K.~Volkov,
``Composite Meson Model with Vector Dominance Based on U(2) Invariant Four Quark Interactions,''
Z. Phys. C \textbf{16} (1983), 205
doi:10.1007/BF01571607


\bibitem{ebert2}
D.~Ebert,
``Bosonization in particle physics,''
Lect. Notes Phys. \textbf{508} (1998), 103-114
doi:10.1007/BFb0106879
[arXiv:hep-ph/9710511 [hep-ph]].



\bibitem{gaillard}
M.~K.~Gaillard,
``The Effective One Loop Lagrangian With Derivative Couplings,''
Nucl. Phys. B \textbf{268} (1986), 669-692
doi:10.1016/0550-3213(86)90264-6

\bibitem{h1}
B.~Henning, X.~Lu and H.~Murayama,
JHEP \textbf{01} (2016), 023
doi:10.1007/JHEP01(2016)023
[arXiv:1412.1837 [hep-ph]].

\bibitem{stergiou}
E.~Mottola, A.~V.~Sadofyev and A.~Stergiou,
``Axions and Superfluidity in Weyl Semimetals,''
[arXiv:2310.08629 [hep-th]].

\bibitem{ellis}
S.~A.~R.~Ellis, J.~Quevillon, P.~N.~H.~Vuong, T.~You and Z.~Zhang,
JHEP \textbf{11} (2020), 078
doi:10.1007/JHEP11(2020)078
[arXiv:2006.16260 [hep-ph]].

\bibitem{ah}
A.~Angelescu and P.~Huang,
``Integrating Out New Fermions at One Loop,''
JHEP \textbf{01} (2021), 049
doi:10.1007/JHEP01(2021)049
[arXiv:2006.16532 [hep-ph]].

\bibitem{smith}
J.~Quevillon, C.~Smith and P.~N.~H.~Vuong,
``Axion effective action,''
JHEP \textbf{08} (2022), 137
doi:10.1007/JHEP08(2022)137
[arXiv:2112.00553 [hep-ph]].





\end{thebibliography}
\end{document}